\newif\ifsubmode
\submodefalse

\ifsubmode
  \documentclass[12pt,preprint]{aastex}  
  \received{}
  \revised{}
  \accepted{}

\else
  \documentclass{emulateapj}  
  \slugcomment{Accepted for Publication in ApJ}
\fi

\shortauthors{Fassnacht et al.}
\shorttitle{Galaxy Groups Associated with B1608+656}

\newcommand{\kms}{km\ s$^{-1}$}
\newcommand{\kmsmpc}{km\ s$^{-1}$\ Mpc$^{-1}$}

\begin{document}

\title{Mass Along the Line of Sight to the
Gravitational Lens B1608+656: Galaxy Groups and Implications for $H_0$
\footnote{\rm Based in
part on observations made with the NASA/ESA Hubble Space Telescope,
obtained at the Space Telescope Science Institute, which is operated
by the Association of Universities for Research in Astronomy, Inc.,
under NASA contract NAS 5-26555. These observations are associated
with program \#GO-10158.}
}

\author{C. D. Fassnacht, R. R. Gal, L. M. Lubin, J. P. McKean}
\affil{
   Department of Physics, University of California, 1 Shields Avenue,
   Davis, CA 95616 }
\email{fassnacht@physics.ucdavis.edu}

\author{Gordon K. Squires}
\affil{
   Spitzer Science Center, California Institute of Technology,
   Mail Code 220-6, 1200 E. California Blvd., Pasadena, CA 91125
}

\and

\author{A. C. S. Readhead}
\affil{
   Astronomy Department, 105-24,
   Caltech,
   Pasadena, CA 91125
}

\begin{abstract}
We report the discovery of four groups of galaxies along the line of
sight to the B1608+656 gravitational lens system.  One group is at the
redshift of the primary lensing galaxy ($z = 0.631$) and appears to
have a low mass, with eight spectroscopically-confirmed members and an
estimated velocity dispersion of 150$\pm$60~\kms.  The three other
groups are in the foreground of the lens.  These groups contain
$\sim$10 confirmed members each, and are located at redshifts of
0.265, 0.426, and 0.520.  Two of the three additional groups are
centered roughly on the lens system, while the third is centered
$\sim$1\arcmin\ south of the lens.  We investigate the effect of each
of the four groups on the gravitational lensing potential of the
B1608+656 system, with a particular focus on the implications for the
value of $H_0$ derived from this system.  We find that each group
provides an external convergence ($\kappa_{ext}$) of
$\sim$0.005--0.060, depending on the assumptions made in calculating
the convergence.  For lens systems where no additional observables
that can break the mass-sheet degeneracy exist, the determination of
$H_0$ will be biased high by a factor of $(1 - \kappa_{ext})^{-1}$ if
the external convergence is not properly included in the model.  For
the B1608+656 system, the stellar velocity dispersion of the lensing
galaxy has been measured, thus breaking the mass-sheet degeneracy due
to the group that is physically associated with the lens.  The effect
of the other groups along the line of sight can be folded into the
overall uncertainties due to large-scale structure (LSS) along the
line of sight.  Because the B1608+656 system appears to lie along an
overdense line of sight, the LSS will cause the measurement of $H_0$
to be biased high for this system.  The systematic bias introduced by
LSS could be 5\% or greater.
Because LSS bias should be random along random lines of sight,
averaging a large number of determinations of $H_0$ from different
lens systems should substantially reduce the uncertainties due to LSS
on the global lens-based measurement of $H_0$.

\end{abstract}


\section{Introduction}

One of the many possible uses of gravitational lens systems is the
measurement of $H_0$ \citep{refsdal}.  However, this method has been
limited by a lack of knowledge about the mass distribution in the
lens, which leads to degeneracies between model parameters and $H_0$.
Perhaps the most important degeneracy is that between the radial slope
of the lensing mass profile and $H_0$.  However, strong limits can be
placed on the mass slope in systems in which measurements of stellar
velocity dispersions can be made \citep[e.g.,][]{tk_galevol}, more
than one component of the source is multiply-imaged
\citep[e.g.,][]{cohn1933}, an Einstein ring is seen
\citep[e.g.,][]{cskrings}, or VLBI structure in the lensed images can
be clearly mapped from one image to another
\citep[e.g.,][]{rusin1152}.  Another major degeneracy is that caused
by an extended mass distribution that is associated with the lens.
This is the famous ``mass-sheet degeneracy''
\citep[e.g.,][]{mass_sheet} and can be caused by a cluster or group
along the line of sight to the main lensing galaxy.  The problem is
that the standard lens observables, namely the locations and fluxes of
the lensed images and the location of the lensing galaxy, do not
indicate how much of the lensing mass surface density is due to the
mass sheet.  In fact, with the exception of very large separation
lenses \citep[e.g., SDSS J1004+4112 and
Q0957+561;][]{1004oguri,0957walsh}, it is difficult to know from the
standard lensing observables whether or not there even exists an
associated group or cluster.  Thus, it is necessary to search for such
structures by other means.  We are conducting a survey of lenses in
which time delays have been measured, with the aim of detecting groups
or clusters which can affect the determination of $H_0$ from those
systems.

The 
\ifsubmode
   \object{CLASS B1608+656}
\else
   CLASS B1608+656
\fi 
system \citep{stm1608} is an excellent target for our survey.  The
redshifts of the lens and background source have been measured to be
$z=0.630$ and $z = 1.39$, respectively \citep{stm1608,zs1608}.  It
remains the only four-image lens system for which robust and
high-precision measurements of all three independent time delays have
been made \citep{1608delays2}.  The lens has been subjected to
intensive modeling, which incorporated information from the measured
stellar velocity distribution and the Einstein ring shape.  With an
advanced modeling code, strong limits were placed on the slope of the
mass density profile, yielding a measurement of $H_0 =
75^{+7}_{-6}$~\kmsmpc\
\citep{1608H0}.  Furthermore, the mass models for this system require
a relatively large external shear of nearly 0.1 \citep{1608H0},
indicating the presence of nearby mass.  In this paper, we provide
evidence for a group of galaxies associated with the primary lensing
galaxy and discuss the result that it and other structures along the
line of sight have on the determination of $H_0$ from this system.

Throughout this paper we assume $\Omega_M = 0.3$, $\Omega_\Lambda = 0.7$,
and, unless otherwise stated, we will express the Hubble Constant as
$H_0 = 100 h~{\rm km}\,{\rm s}^{-1}\,{\rm Mpc}^{-1}$.

\section{Observations and Data Reduction}

We have conducted a spectroscopic survey of the B1608+656 field as
part of our ongoing program to find compact groups of galaxies
associated with gravitational lenses.  The field was imaged in three
bands, Gunn $g$, $r$, and $i$ \citep{gunnstd}, with the Palomar
60-Inch Telescope.  Spectroscopic targets were selected based on their
colors and distances from the lens system.  The highest priority
targets were those close to the lens system and with $(r - i)$ colors
in the range 0.45 to 0.65, i.e., close to those expected for
early-type galaxies at the redshift of the lensing galaxy.  All of the
spectroscopic targets had $r \leq 23$.  In addition to the high
priority targets, several other galaxies were observed in order to
pack efficiently the slitmasks that were used for the bulk of the
spectroscopy.  The spectroscopic observations were made with the Low
Resolution Imaging Spectrograph \citep[LRIS;][]{lris}, in both
longslit and multislit modes, and the Echellete Spectrograph and
Imager \citep[ESI;][]{esi} on the W. M. Keck Telescopes.
The spectroscopic data were reduced using scripts based on standard
{\sc iraf}\footnote{IRAF (Image Reduction and Analysis Facility) is
distributed by the National Optical Astronomy Observatories, which are
operated by the Association of Universities for Research in Astronomy
under cooperative agreement with the National Science Foundation.}
tasks.  Additional IDL scripts were used to process the ESI data.
The spectroscopic and photometric data on all of the surveyed
galaxies, along with the full details of the data reduction
procedures, will be presented in a future paper dealing with the 
properties of the groups and the galaxies within them.

\section{Mass Along the Line of Sight to B1608+656}

\begin{figure}
  \plotone{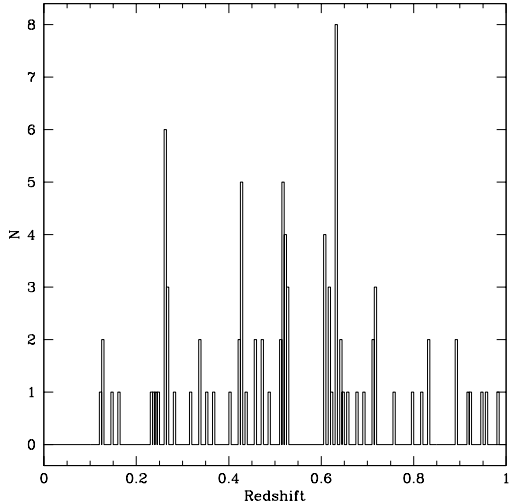}
  \caption{
   Histogram showing the distribution of the 97 non-stellar redshifts 
   obtained in the field of B1608+656.  The width of the bins is
   $\Delta z = 0.005$.  The most prominent spike, with eight galaxies,
   is at the redshift of
   the lensing galaxy ($z = 0.63$).
   \label{fig_zhist}}
  \end{figure}

The spectroscopic observations produced redshifts for 97 galaxies in
the field.  The typical uncertainties in the redshifts, based on the
scatter of redshifts calculated from individual lines in each
spectrum, were $\sigma_z \sim 0.0003$.  The distribution of redshifts,
in which several spikes are seen, is shown in Figure~\ref{fig_zhist}.
We select group or cluster candidates by looking for structures that
are concentrated both spatially and in redshift space, using
the iterative method described by \citet{wilmangrp1}.  We investigated
structures with at least five members in one redshift bin, where
the bin size was $\Delta z = 0.005$.  

To characterize each group, we estimate its line-of-sight velocity
dispersion ($\sigma_v$) utilizing the ROSTAT package (Beers, Flynn, \&
Gebhardt 1990).  This package avoids assumptions of gaussianity in the
underlying velocity distributions and includes bootstrap and jackknife
error estimation for the group redshifts and dispersions. Many of
these estimators are significantly more resistant to non-gaussianity
in the sample than the traditional Gaussian; large differences between
the different scale measures can be a sign of deviations from
gaussianity in the sample, either in shape or due to outliers. In
particular, we use three estimators.  The first, $\sigma_{gap}$, is
based on the gapper algorithm, which is recommended for very small
datasets \citep{bee90}.  The error on this estimator is the jackknifed
biweight estimate.  The second is the biweight scale estimate,
$\sigma_{biwt}$, which is recommended for datasets with $\sim10$
members.  Its confidence interval is taken from the jackknifed gapper.
Finally, we use $\sigma_{gauss}$, the Gaussian estimator, and its
1-sigma Student's $t$ error.  Consistency among the various dispersion
estimates for a small redshift sample lends credence to the
measurement.  However, we note that velocity dispersions based on a
small number of redshifts can be highly
uncertain \cite[e.g.,][]{zm98}.

\subsection{The Group Associated with B1608+656}

The most prominent spike in the redshift distribution consists of
eight galaxies with redshifts of $z \sim 0.63$.  This spike includes
the lensing galaxy at $z_\ell = 0.630$.  However, we have not included
the second lensing galaxy within the ring of images
\citep[G2;][]{1608H0,1608surpi}.  All indications are that G2 is
merging with G1, and therefore that this group consists of at least
nine members.  As one test of the likelihood that this redshift spike
represents a real group, we have plotted the spatial distribution of
the galaxies in the spike (Figure~2).  We find that the galaxies are
spatially concentrated and centered roughly on the position of the
lens.  Seven of the eight galaxies in the redshift spike, or eight out
of nine if G2 is included, are within a circle centered on the lens
system of radius 2\farcm1, corresponding to 1~$h^{-1}$ comoving Mpc at
the redshift of the lens.  Figure 3 shows \ the redshift distribution
in the region of the spike in terms of velocity offsets from the mean
redshift of the group.  The galaxies also have a tight distribution in
velocity space, with all eight within $\pm$300~\kms\ of the mean
redshift.  Thus, we conclude that these galaxies represent a group,
hereafter called Group 1, associated with the lensing galaxy.

For Group 1, we find that the various $\sigma_v$ estimators obtained
from the ROSTAT package are all consistent. This result strongly
suggests that, despite the small number of redshifts in the group, the
velocity distribution is well described by a Gaussian, and the
resulting velocity dispersion is robust.  In Table~\ref{tab_groupdata}
we list the median redshift and the three estimates of the
line-of-sight velocity dispersion ($\sigma_{gap}$, $\sigma_{biwt}$,
and $\sigma_{gauss}$) with their associated errors (in \kms).
Based on the number of group members, we use $\sigma_v = \sigma_{gap}$ 
for Group 1, since $\sigma_{gap}$ is the more appropriate estimator
for very small data sets \citep{bee90}.  Thus, we obtain $\sigma_v =
150 \pm 60$~\kms.

\ifsubmode
  \clearpage
\fi

  \begin{figure}
  \plotone{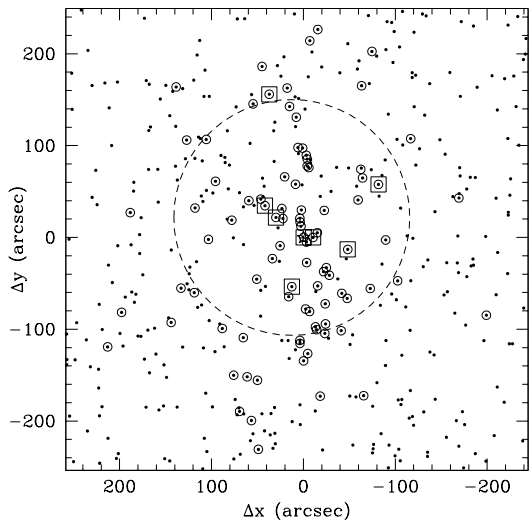}
  \caption{
   Spatial distribution of the galaxies in Group 1.  The field of view
   is 10\arcmin x 11\arcmin, with the axes labeled in terms of offsets from the
   B1608+656 lens system in arcseconds.  
   The dots represent the positions of galaxies with
   $r \leq 23$, while the open circles mark the galaxies for which redshifts
   have been obtained.    
   The open boxes mark the galaxies in the group.
   The large dashed circle has a radius of 1 $h^{-1}$ comoving Mpc at
   the redshift of the lensing galaxy.
   \label{fig_group1_zgals}}
  \end{figure}

  \begin{figure}
  \plotone{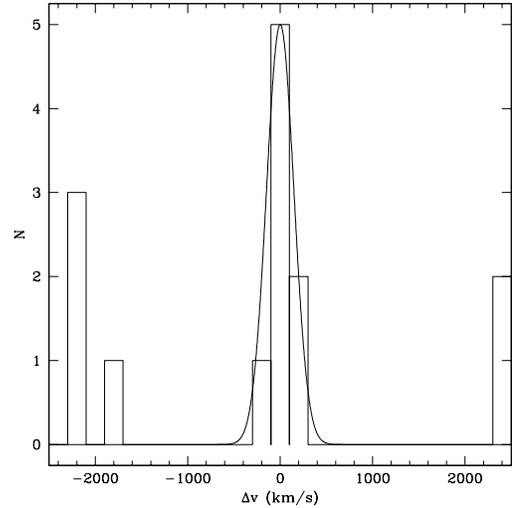}
  \caption{
   Histogram showing the velocity distribution of the galaxies in and
   surrounding the redshift spike at $z = 0.63$.  The bin widths are 
   200~\kms, approximately twice the uncertainties in determining
   the redshifts.  The curve is a Gaussian with $\sigma = $150~\kms.
   \label{fig_vhist1608}}
  \end{figure}
\ifsubmode
   \clearpage
\fi

\subsection{Other Mass Concentrations Along the Line of Sight}

There are other spikes in the distribution of measured redshifts shown
in Figure~\ref{fig_zhist}.  In addition to the group at the redshift
of the lens, there are three other group candidates in the observed
distribution satisfying the redshift and spatial concentration
criteria, with mean redshifts of $<z> = 0.265$ (Group 2), $<z> =
0.426$ (Group 3), and $<z> = 0.520$ (Group 4).  Each group has a
substantial number of confirmed members, with sizes of nine, seven,
and 14 galaxies, respectively.  The properties of these groups are
given in Table~\ref{tab_groupdata}.  For each group, we find that the
various $\sigma_v$ estimators are all consistent.  Based on the number
of confirmed members in each group, we estimate the group velocity
dispersions with $\sigma_{gap}$ for Groups 2 and 3, and
$\sigma_{biwt}$ for Group 4.  Figures \ref{fig_xy} and
\ref{fig_vhist} show the spatial and velocity distributions,
respectively, for the three additional groups detected in this field.
We explore the effects that these groups may have on the determination
of $H_0$ from B1608+656 in the following sections.
\ifsubmode
  \clearpage
  \begin{figure}
  \plotone{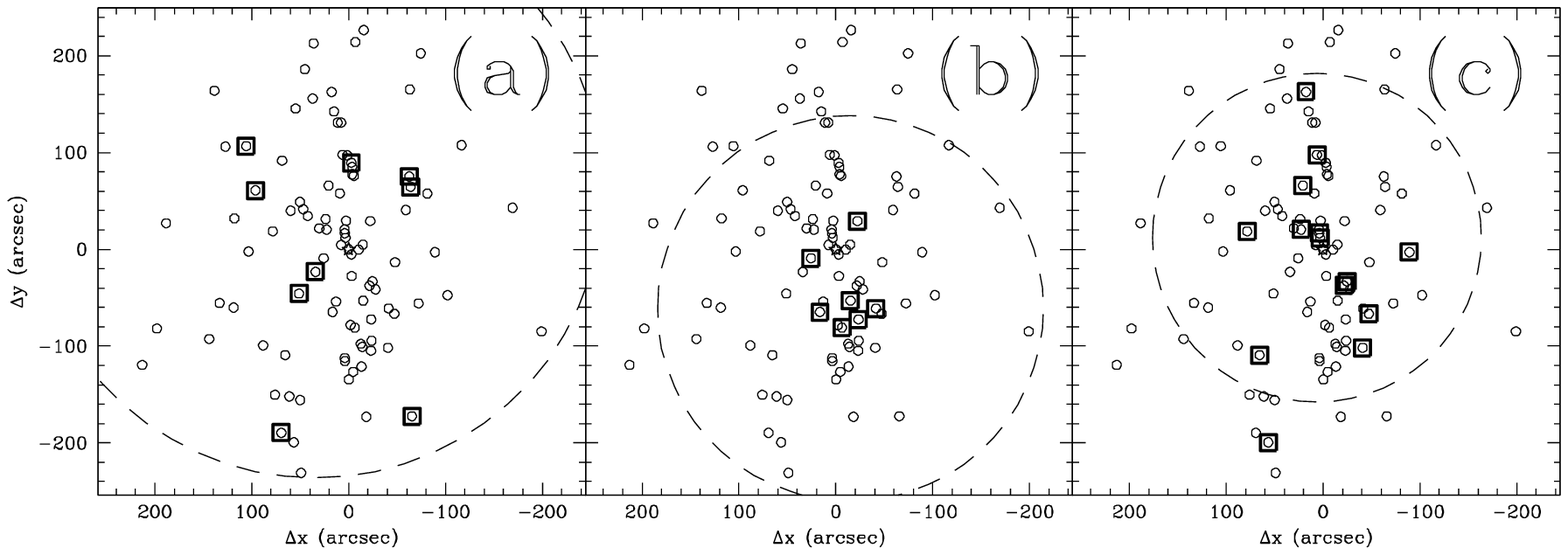}
  \caption{
   Spatial distributions of the galaxies in the additional three group
   candidates in the B1608+656 field, represented as in 
   Figure~\ref{fig_group1_zgals}.  The open circles represent galaxies
   for which redshifts have been obtained, while the open boxes 
   represent confirmed 
   group members.  The dashed circle in each plot
   has a radius of 1$h^{-1}$ comoving Mpc at the redshift
   of the group and is centered at the median position of
   the confirmed group members.  
   (a) Group 2 at $z \sim 0.27$.  (b) Group 3 at $z \sim 0.43$.  
   (c) Group 4 at $z \sim 0.52$.
   \label{fig_xy}}
  \end{figure}
\else
  \begin{figure*}
  \plotone{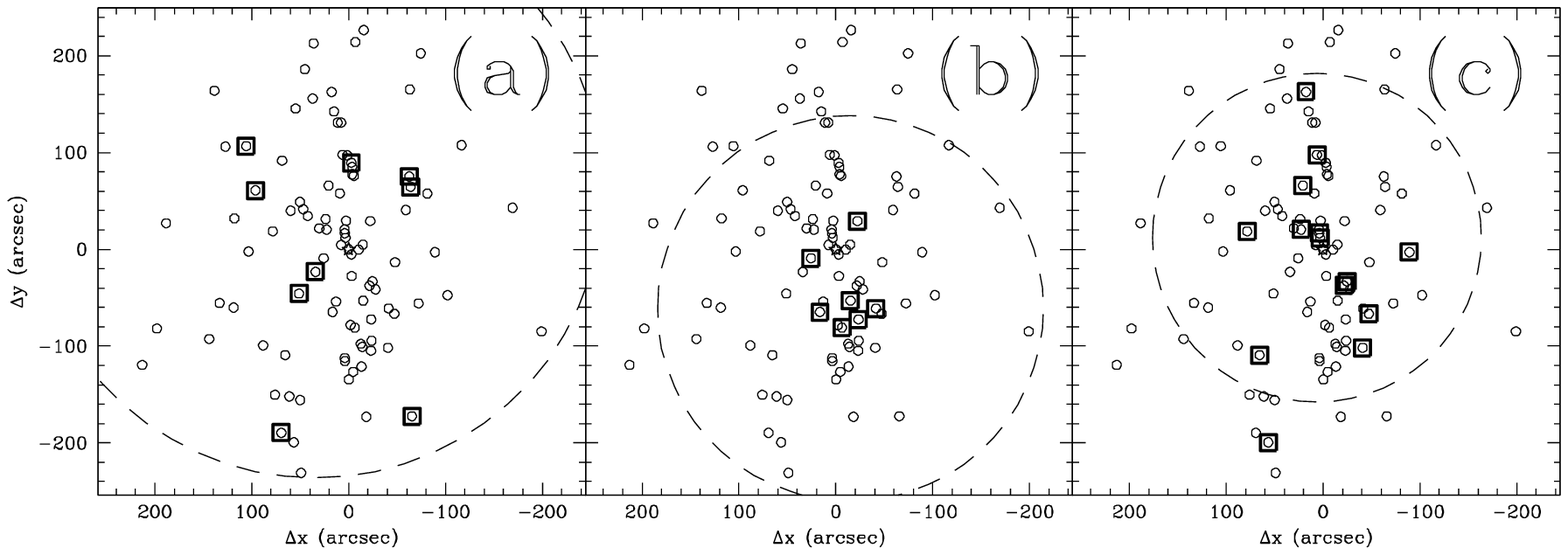}
  \caption{
   Spatial distributions of the galaxies in the additional three group
   candidates in the B1608+656 field, represented as in 
   Figure~\ref{fig_group1_zgals}.  The open circles represent galaxies
   for which redshifts have been obtained, while the open boxes 
   represent confirmed 
   group members.  The dashed circle in each plot
   has a radius of 1$h^{-1}$ comoving Mpc at the redshift
   of the group and is centered at the median position of
   the confirmed group members.  
   (a) Group 2 at $z \sim 0.27$.  (b) Group 3 at $z \sim 0.43$.  
   (c) Group 4 at $z \sim 0.52$.
   \label{fig_xy}}
  \end{figure*}
\fi

\begin{center}
\ifsubmode
   \clearpage
   \begin{deluxetable}{cclrlllrrrr}
\else
   \begin{deluxetable*}{cclrlllrrrr}
\fi
\tabletypesize{\scriptsize}
\tablecolumns{10}
\tablewidth{0pc}
\tablecaption{Group Properties}
\tablehead{
\colhead{}
 & \colhead{}
 & \colhead{}
 & \colhead{}
 & \colhead{$\sigma_{gap}$}
 & \colhead{$\sigma_{biwt}$}
 & \colhead{$\sigma_{gauss}$}
 & \colhead{$\theta_{med}$\tablenotemark{a}}
 & \colhead{PA$_{med}$\tablenotemark{a}}
 & \colhead{$\theta_{lw}$\tablenotemark{b}}
 & \colhead{PA$_{lw}$\tablenotemark{b}} \\
\colhead{Group}
 & \colhead{$<z>$}
 & \colhead{$D_{\ell s}/D_s$}
 & \colhead{$N_{gals}$}
 & \colhead{(km s$^{-1}$)}
 & \colhead{(km s$^{-1}$)}
 & \colhead{(km s$^{-1}$)}
 & \colhead{(arcsec)}
 & \colhead{($\degr$)}
 & \colhead{(arcsec)}
 & \colhead{($\degr$)}
}
\startdata
1 & 0.6313 & 0.45 &  8\tablenotemark{c} & 150$\pm$60  & $130\pm100$ & $150^{+50}_{-20}$ &
   25 & 30   &  8 &  $-$5 \\
2 & 0.2651 & 0.74 &  9 & 320$\pm$100 & $230\pm270$ & $320^{+120}_{-100}$ &
   69 & 29   & 18 &   6 \\
3 & 0.4263 & 0.66 &  7 &  270$\pm$110 & $260\pm120$ & $270\pm110$ &
   63 & -166 & 46 & $-$155 \\
4 & 0.5202 & 0.53 & 14 & 920$\pm$120 & $930\pm140$ & $890^{+240}_{-130}$ &
   13 & 26   &  6 &  $-$47 \\
4a\tablenotemark{d} & 0.5163 & 0.53 & 7 & 410$\pm$160 & 100$\pm$320 & 430$\pm$110 &
   24 & 47   & 46 &  6 \\
4b\tablenotemark{d}  & 0.5241 & 0.53 & 7 & 360$\pm$120 & 350$\pm$120 & 350$\pm$120 &
   44 & $-$150 & 52 &  $-$161 \\
\enddata
\tablenotetext{a}{Median group position expressed as an offset from the lens.
The PA is measured north through east.}
\tablenotetext{b}{The luminosity-weighted group position expressed as an offset
from the lens. The PA is measured north through east.}
\tablenotetext{c}{This number does not include the second galaxy (G2) inside the
Einstein ring of the lens system because there is no measured redshift for G2.
However, G2 appears to be merging
with the primary lens galaxy and thus appears to be a member of the group.}
\tablenotetext{d}{Groups 4a and 4b are subsets of Group 4.  See \S4.1.}
\label{tab_groupdata}
\ifsubmode
   \end{deluxetable}
\else
   \end{deluxetable*}
\fi
\end{center}

\ifsubmode
  \clearpage
  \begin{figure}
  \plotone{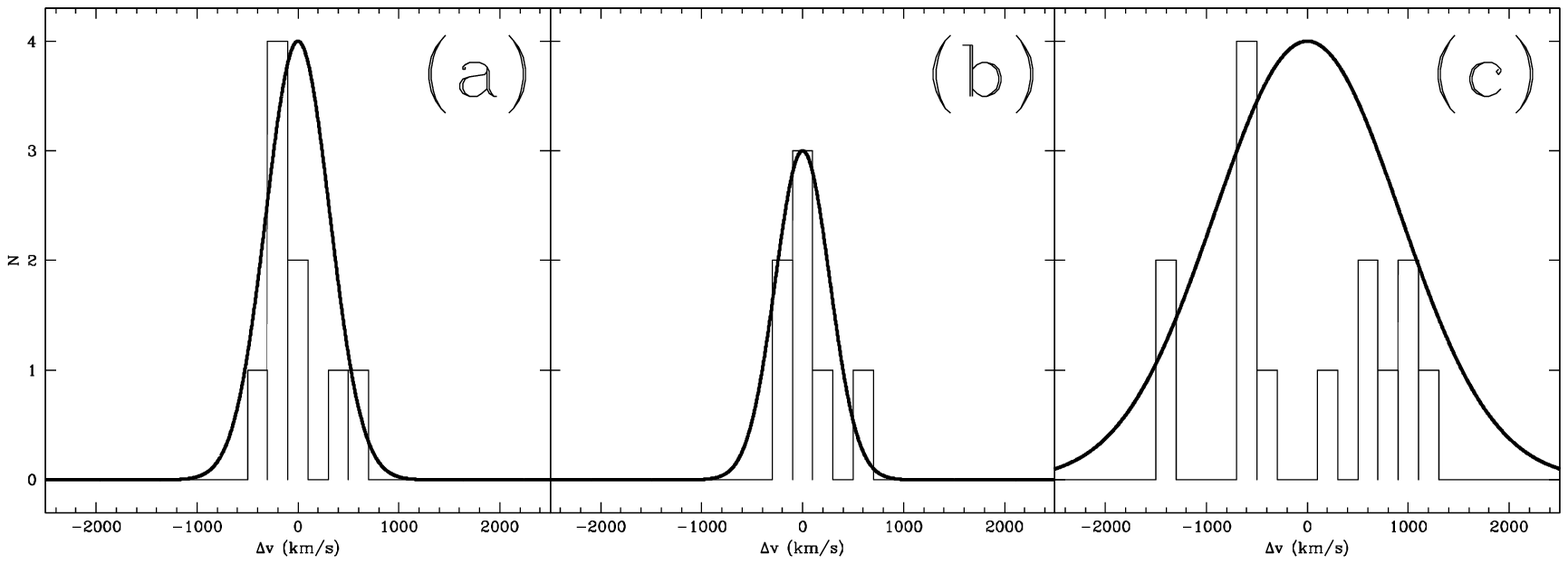}
  \caption{
   Histograms showing the velocity distributions of the galaxies in the three
   other spatially-concentrated redshift spikes.  The bin widths are 
   200~km\,s$^{-1}$, approximately twice the uncertainties in determining
   the redshifts.
   \label{fig_vhist}}
  \end{figure}
  \clearpage
\else
  \begin{figure*}
  \plotone{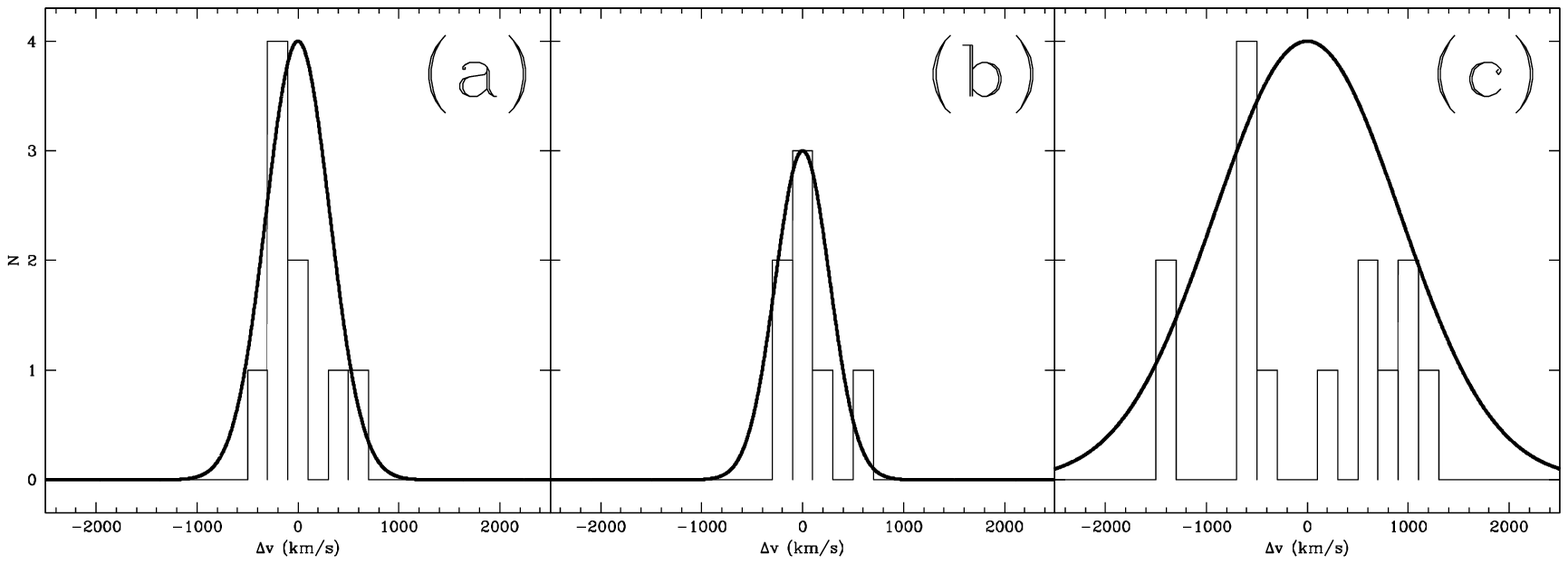}
  \caption{
   Histograms showing the velocity distributions of the galaxies in the three
   other spatially-concentrated redshift spikes.  The bin widths are 
   200~km\,s$^{-1}$, approximately twice the uncertainties in determining
   the redshifts.
   \label{fig_vhist}}
  \end{figure*}
\fi

\section{Discussion}

\subsection{A Cluster at $z = 0.52$?}

Of the three additional candidate groups, the one at $z = 0.520$ stands
out.  Its velocity dispersion of $\sim$900~\kms\ implies that this
structure is a cluster of galaxies rather than a group.  If it is in
fact a real cluster, it will have a significant effect on the measured
value of $H_0$ obtained from the B1608+656 lens system (see next
section).  Therefore, we consider evidence that pertains to the
reality of the cluster.

\begin{figure}
  \plotone{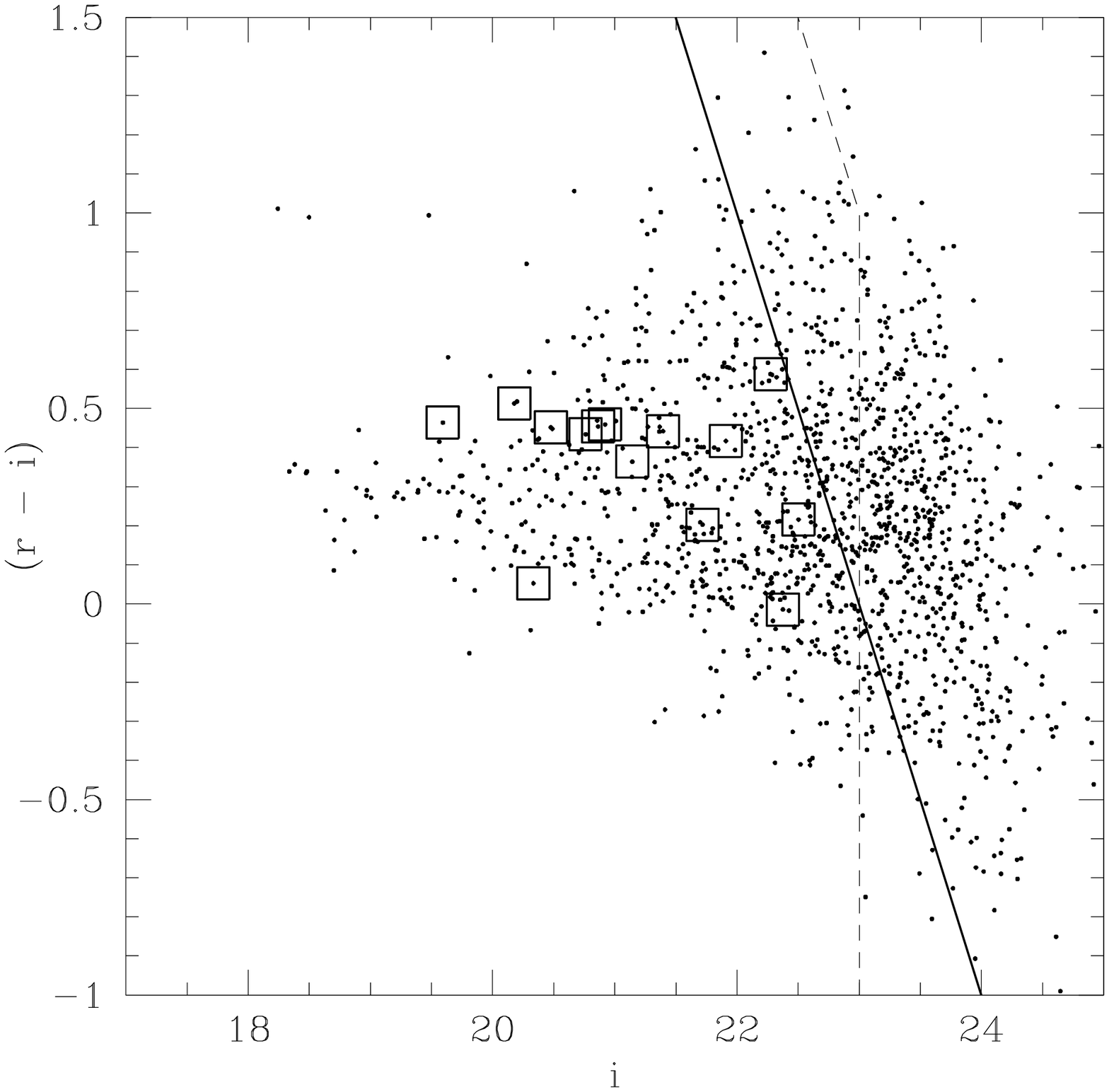}
  \caption{
   Color-magnitude diagram for the B1608+656 field.  The dashed lines
   represent the approximate completeness limit of the imaging obtained
   at the Palomar 60-Inch Telescope.
   The boxed points belong to the $z=0.52$ cluster candidate.  
   The diagonal solid line represents the limit of the spectroscopy at
   $r \sim 23$.
   \label{fig_cmri}}
  \end{figure}

There are two main arguments in favor of the redshift spike's
representing a real cluster.  The first is the concentration of the
structure in both velocity space and in projection on the sky.  Nine
of the 14 spectroscopically-confirmed galaxies in the structure are
within $\pm$1000\kms\ of the mean redshift, while 13 of 14 lie within
a projected distance of 1 $h^{-1}$ comoving Mpc of the median
position.  The second is that nine of the 14 galaxies in the group lie
along a tight sequence on a color-magnitude diagram, at $(r - i) \sim
0.45$ (Figure \ref{fig_cmri}).

The arguments against the reality of the cluster are as follows.
First, an examination of deep Advanced Camera for Surveys (ACS) images
of this field (Program GO-10158; PI: Fassnacht) shows no obvious
overdensity such as might be expected from a cluster.  Furthermore, of
the 11 galaxies that are members of Group 4 and also lie within the
field of view of the ACS imaging, none appears to be a central bright
early-type galaxy.  In fact, nine out of the 11 group galaxies in the
ACS imaging appear to be spirals.  This is surprising given the red
colors of these galaxies, but we will leave the discussion of this
point to the followup paper in which we will focus on the properties
of the groups and the galaxies within them.  The second point arguing
against the cluster hypothesis is that the group velocities (Figure
5c) are not normally distributed as one would expect from a relaxed
cluster.  With the large velocity dispersion implied by the measured
redshifts, it is perhaps not unlikely that a random set of 14 galaxies
does not produce a nice Gaussian shape.  Further spectroscopy of the
field, especially pushing deeper than the $r\sim 23$ limit used in the
previous observations, may fill in the velocity structure and produce
a velocity distribution that is closer to Gaussian.  On the other
hand, the last slitmask in our observing program had the targets
re-prioritized to preferentially find galaxies in the cluster.  Only
one additional member was found, once again indicating that there is
not the large overdensity of galaxies typically found in a cluster.
Third, there was no extended X-ray emission detected in this field in
a recent 30~ksec Chandra X-ray Observatory observation of this field
\citep{chandra1608}.  If the cluster is as massive as suggested by its
velocity dispersion and is virialized, it should have been detected by
the Chandra observations, which had a 3-$\sigma$ upper limit for
detection that correspond to a velocity dispersion of $\sim$500~\kms\
\citep{chandra1608}.  However, it is possible that the cluster is
recently formed or is, in fact, a pair of merging groups.  In these
situations, it may be severely underluminous at X-ray wavelengths.
Finally, a cluster this massive and centered as close to the lens as
it appears to be would introduce a strong effect on the lensing
signature.  The external shear introduced by such a cluster
would be much larger than required by the lens model \citep{1608H0}.
Furthermore, it is possible that the surface mass density would be
high enough at the position of the lens to form yet another image of
the lensed source, i.e., the cluster would have to be treated in the
strong lensing regime.

\ifsubmode
  \clearpage
\fi
\ifsubmode
  \clearpage
\fi

We feel that the arguments against the reality of the cluster are
stronger than those in favor.  The structure may instead be a pair of
merging groups or some kind of filamentary structure.  To explore the
two-group hypothesis, we arbitrarily split the group into two parts by
redshift.  Group 4a is defined as the seven galaxies with redshifts
smaller than the median Group 4 redshift, while Group 4b is comprised
of the seven galaxies with redshifts larger than the median.
Figure~\ref{fig_2group} shows the spatial and velocity distributions
of the two groups.  The velocity dispersions of the two groups are
$\sim 400$~\kms\ and $\sim 350$~\kms.  If the two-group explanation is
correct, and these velocity dispersions are close to the true values,
then it is not surprising that the X-ray observations of
\citet{chandra1608} did not detect diffuse X-ray emission.  The
spatial distribution of Group 4b is centered slightly to the southwest
of the Group 4a centroid.  However, the two centroids are completely
consistent, given the uncertainties.  Another possibility is that the
Group 4 galaxies lie along a filament.  We note that much of the
following discussion treats each group as a collection of individual
halos, each associated with a group galaxy.  Therefore, whether the
galaxy under discussion is in a cluster, a group, or a filament is
irrelevant.

\ifsubmode
  \clearpage
  \begin{figure}
  \plotone{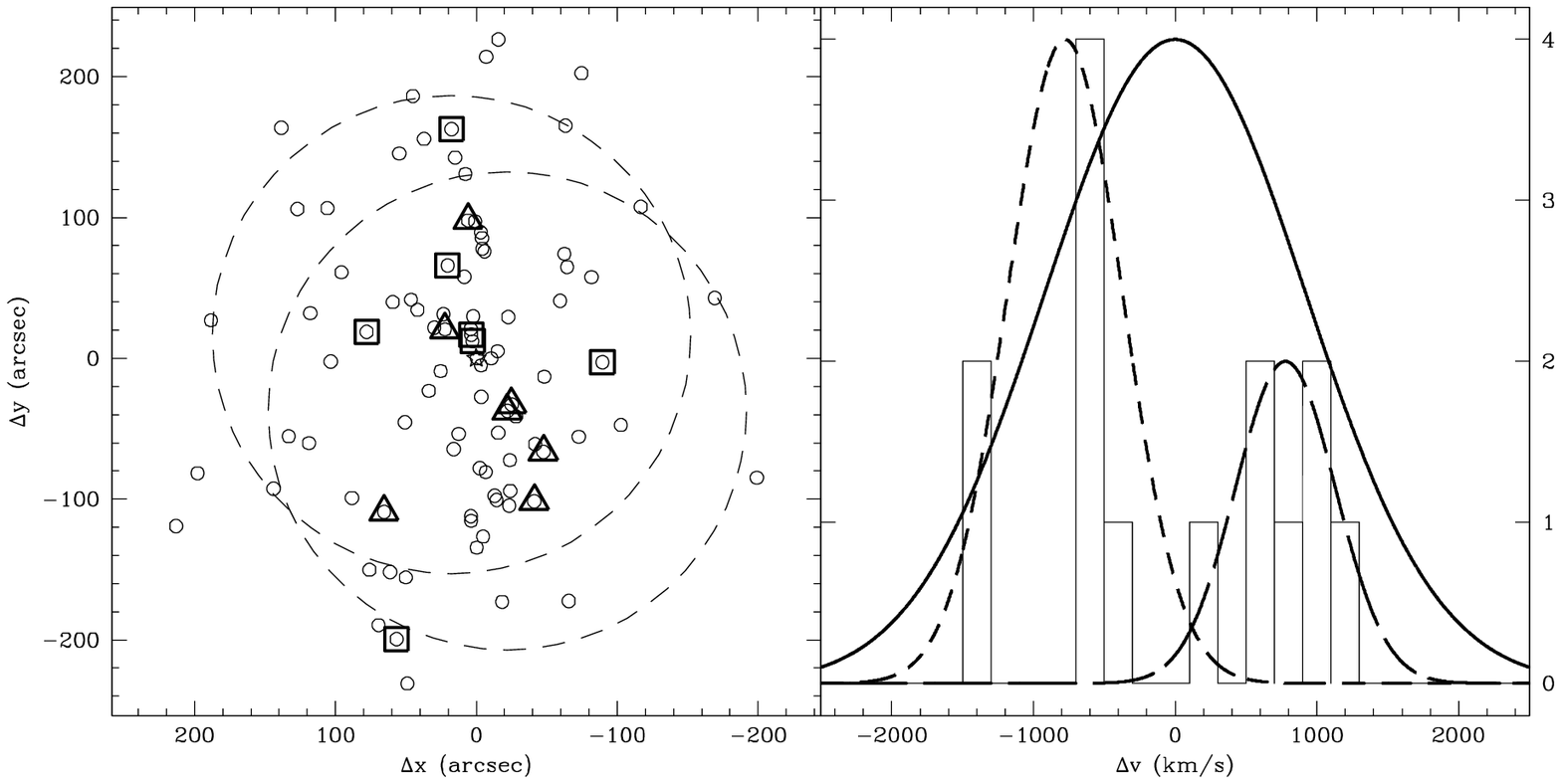}
  \caption{
   Spatial (left) and velocity (right) distributions of galaxies in 
   Group 4.  The galaxies have been split into two smaller groups
   in velocity space.  The spatial distribution of Group 4a (redshifts
   smaller than the median Group 4 redshift) are marked by boxes.
   The galaxies in Group 4b are marked with triangles.
   \label{fig_2group}}
  \end{figure}
  \clearpage
\else
  \begin{figure*}
  \plotone{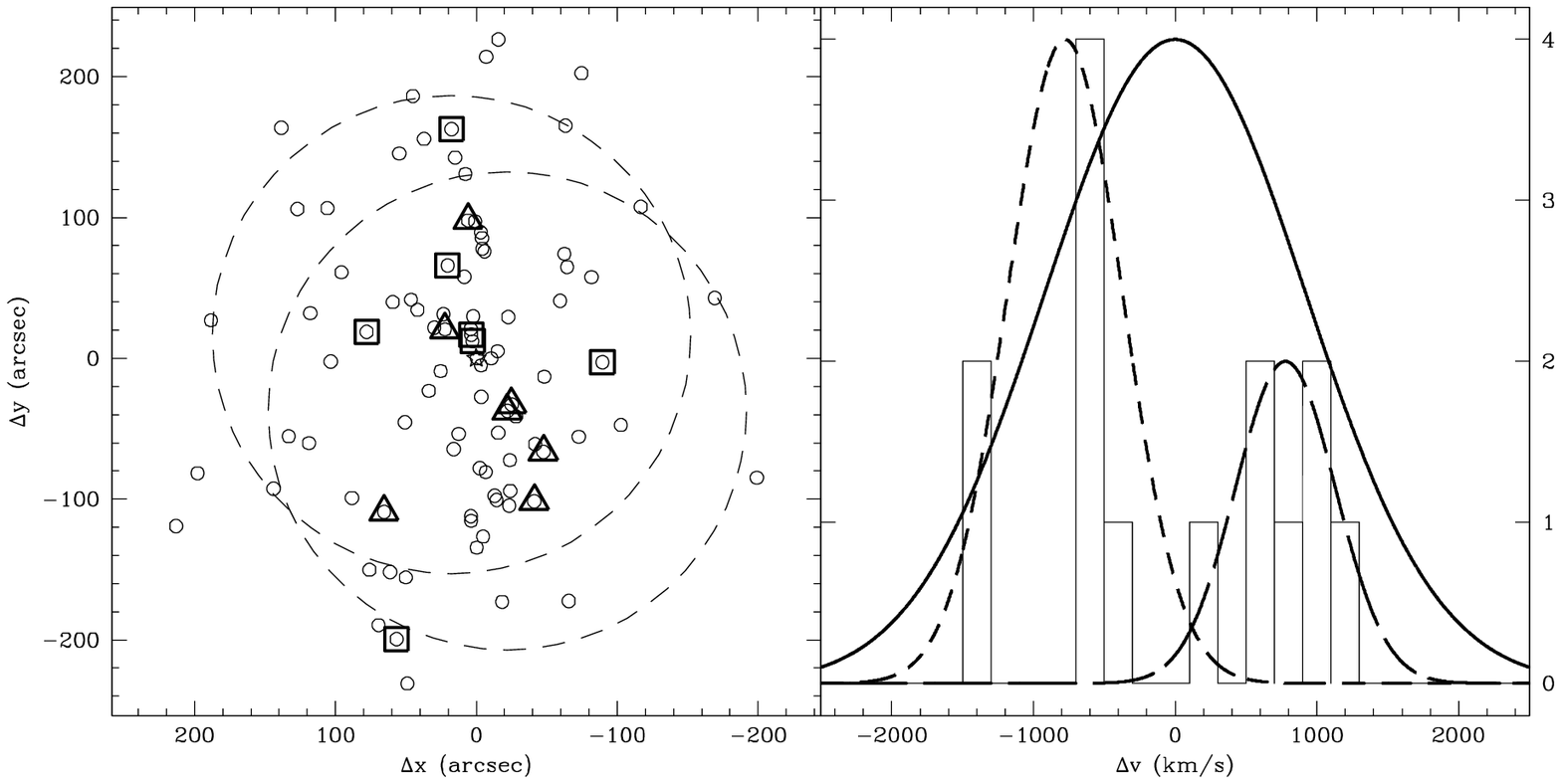}
  \caption{
   Spatial (left) and velocity (right) distributions of galaxies in 
   Group 4.  The galaxies have been split into two smaller groups
   in velocity space.  The spatial distribution of Group 4a (redshifts
   smaller than the median Group 4 redshift) are marked by boxes.
   The galaxies in Group 4b are marked with triangles.
   \label{fig_2group}}
  \end{figure*}
\fi

\subsection{Effect on Gravitational Lensing\label{sec_lensing}}

We now consider the effect the groups associated with B1608+656 may
have on the overall lensing gravitational potential, with a particular
emphasis on the effect on $H_0$.  While the angular separation of the
lensed images in a gravitational lens system provides an accurate
measurement of the projected lensing mass, it does not require that
all of the mass be associated with the primary lensing galaxy.  In
fact, some of the mass can be contributed by other structures along
the line of sight, such as the groups discussed above.  The
contribution of external mass is quantified through the convergence,
$\kappa_{ext}$, that it causes at the location of the lensing galaxy.
The convergence is just the scaled mass surface density:
$$
\kappa_{ext} = \frac{\Sigma_{ext}}{\Sigma_c}; \quad
\Sigma_c = \frac{c^2}{4 \pi\ G} \frac{D_s}{D_\ell\ D_{\ell s}}, 
$$
where $\Sigma_c$ represents the critical mass surface density required
for multiple images to form.  As usual, $D_\ell$, $D_{\ell s}$, and
$D_s$ are the angular diameter distances between observer and lens,
lens and source, and observer and source, respectively.  The
external convergence will lead to a value of $H_0$ that is too high
if the full lensing mass is improperly assigned solely to the primary
lensing galaxy, i.e.,
$$
H_{0,true} = H_{0,meas} (1 - \kappa_{ext}),
$$
where $H_{0,meas}$ is the value obtained without properly including
the external convergence in the lens model.  We will use two methods
to estimate the external convergence contributed by each of the groups
along the line of sight to the B1608+656 system.  Although neither of
these methods may be entirely correct, the range of results that
they produce should be representative of the true group convergences.

The first, and more traditional, method is to assume that the group
can be approximated as a smooth mass distribution.  The distribution,
for simplicity, is usually taken to be that produced by a singular
isothermal sphere (SIS).  
In this case, the convergence contributed by
the group, calculated at the location of the lens, is
$$
\kappa_{SIS}(\theta_{cent}) = 
 \frac{D_{\ell s}}{D_s} \frac{2 \pi \sigma_v^2}{c^2 \theta_{cent}}
 = \frac{b_{SIS}}{2 \theta_{cent}},
$$
where $\theta_{cent}$ is the angular offset between the center of the group
and the lens system.  The ``lens strength'' of an isothermal
distribution is defined as $b = 4 \pi \sigma_v^2 D_{\ell s} / (D_s
c^2)$, and for a singular isothermal sphere gives the Einstein ring
radius in angular units.  If $\theta_{cent}$ is measured in arcminutes and
$\sigma_v$ is measured in \kms, then
$$
\kappa_{SIS}(\theta_{cent}) \sim 
0.015\ \left ( \frac{D_{\ell s}}{D_s} \right )
\left ( \frac{\sigma_v}{250~{\rm km\ s}^{-1}} \right )^2
\left ( \frac{\theta_{cent}}{{\rm arcmin}} \right )^{-1}.
$$
The convergences due to the cluster and group candidates, calculated
using the SIS assumption, are given in Table~\ref{tab_groupconv}.  In
each case the velocity dispersion used was that obtained from the
gapper method.
With the strong dependence of $\kappa_{ext}$ on $\theta_{cent}$, it
becomes imperative to locate the group centroid accurately, which is
extremely difficult with fewer than several tens of confirmed group
members.  For example, the centroid estimates listed in
Table~\ref{tab_groupdata} differ substantially depending on whether
they were obtained by taking the median position or the
luminosity-weighted position.  
Therefore, we also apply a second
method for estimating the group convergence.

\begin{center}
\ifsubmode
   \clearpage
   \begin{deluxetable}{cllccc}
\else
   \begin{deluxetable*}{cllccc}
\fi
\tabletypesize{\scriptsize}
\tablecolumns{10}
\tablewidth{0pc}
\tablecaption{Convergences due to Groups}
\tablehead{
\colhead{Group}
 & \colhead{$\kappa_{SIS,med}$\tablenotemark{a}}
 & \colhead{$\kappa_{SIS,lw}$\tablenotemark{b}}
 & \colhead{$\kappa_{ind}$\tablenotemark{c}}
 & \colhead{$\kappa_{trunc}$\tablenotemark{c}}
 & \colhead{$N_{trunc}$\tablenotemark{d}}
}
\startdata
1  & 0.0056  & 0.018   & 0.025        & 0.012          & 1 \\
2  & 0.016   & 0.061   & 0.013--0.026 & 0.0040--0.0082 & 1 \\
3  & 0.011   & 0.015   & 0.014--0.028 & 0.0064--0.013  & 2 \\
4  & \nodata & \nodata & 0.026--0.053 & 0.015--0.031   & 3 \\
4a & 0.054   & 0.028   & \nodata      & \nodata        & \nodata \\
4b & 0.023   & 0.019   & \nodata      & \nodata        & \nodata \\
\enddata
\tablenotetext{a}{Convergence calculated with the group represented as
a SIS and the group centroid represented by the median galaxy position.}
\tablenotetext{b}{Convergence calculated with the group represented as
a SIS and the group centroid represented by the luminosity-weighted
mean.}
\tablenotetext{c}{Range corresponds to a range of stellar velocity
dispersions for the fiducial (brightest) galaxy in Groups 2, 3, and 4.
The velocity dispersions range from 140--200~\kms.}
\tablenotetext{d}{Number of galaxies contributing to $\kappa_{trunc}$.}
\label{tab_groupconv}
\ifsubmode
   \end{deluxetable}
   \clearpage
\else
   \end{deluxetable*}
\fi
\end{center}

The alternate method for computing the group convergence is to treat
the group as a collection of individual galaxy halos, with no overall
group halo.  In other words, the mass sheet at the position of the
lensing galaxy is composed of the overlapping halos of the other
galaxies in the group.  The use of this method is motivated by two
considerations.  First, by ignoring any shared group halo, this method
explores what is probably an extreme case, especially when assuming
that the galaxy halos may be truncated (see below).  This extreme case
should provide a fairly robust lower limit to the estimate of the
group convergence.  Second, these moderate-redshift groups may still
be in the early phases of formation and thus the galaxies may not yet
have lost a significant fraction of their individual halos to the
shared group halo.  Simulations performed by \citet{kzgroups} indicate
that similar results are obtained whether the groups are treated as a
single shared halo or a collection of individual halos.  Furthermore,
the input cosmological parameters (e.g., $H_0$) are recovered
accurately when the group contribution is treated as a collection of
galaxy halos.  This accuracy is obtained even though they make the
simplifying assumption that the group galaxies are circular when, in
fact, their simulated galaxy mass distributions were elliptical.  We
follow the approach of
\citet{kzgroups} and assign group galaxy masses (expressed in terms of
their lens strengths, $b_i$) based on their optical magnitudes ($m_i$)
via the relationship
$$
b_i = b_{fid}\ 10^{-0.2(m_i - m_{fid})},
$$
where $b_{fid}$ and $m_{fid}$ are the lens strength and magnitude for
a fiducial galaxy within the group.  In each case, we take the
fiducial galaxy to be the brightest confirmed group member.  If each
of the group members is treated as a singular isothermal sphere, as
in \citet{kzgroups}, then the total convergence at the position of the
lens is just the sum of the individual convergences at that position:
$$
\kappa_{ind} = \sum_i \kappa_i = \sum_i \frac{b_i}{2 \theta_i},
$$  
where $\theta_i$ is the distance between galaxy $i$
and the lens.  The results for each of the four groups are given
in Table~\ref{tab_groupconv}, while notes on the individual groups
are given below.

Another effect to consider is that due to the truncation of the dark
matter halos of the group galaxies. In the calculation of
$\kappa_{ind}$, all of the galaxy halos are assumed to be larger than
the separation between the galaxies and the primary lens system.
However, weak lensing studies by \citet{halo_cutoff} have suggested
that galaxy halos have a truncation radius of $\sim 200 h^{-1}$~kpc.
If this truncation radius is real and is typical, then the convergence
due to galaxies located farther from the lens than this will drop off
faster than the $(1/\theta)$ assumed in the calculation of
$\kappa_{ind}$.  To explore the possible size of this effect, we
recalculate the external convergence contributed by each group,
summing only the contributions from the galaxies that lie within a
projected distance of $200 h^{-1}$ comoving kpc from the lens.  The
results are given as $\kappa_{trunc}$ in Table~\ref{tab_groupconv}.
This is an extreme approach, but it should give an approximate lower
limit to the convergence caused by each group.

\subsubsection{Convergence Due to Group 1}

For this group we use F814W magnitudes measured from {\em Hubble Space
Telescope} observations of the field.  These observations consist of
the deep ACS images mentioned above, as well as Wide-Field Planetary
Camera 2 images of the same field (Program GO-6555: PI Schechter).
The fiducial galaxy is the primary lensing galaxy (G1) which has a
F814W magnitude of 18.2 and a lens strength of $b_{1608} = 0\farcs83$
\citep{1608H0}.  The group galaxy most distant from the lens system is
not covered by the HST images so we do not include it in the
calculation of $\kappa_{ind}$.  However, it is far enough away
(2\farcm7) that its contribution to the overall convergence is
negligible.

\subsubsection{Convergence Due to Group 4}

As discussed in \S4.1, we do not believe that Group 4 is a real
cluster.  We note that an assumption that the cluster is real yields
extremely large convergences at the position of the B1608+656 system.
The SIS approximation leads to $\kappa_{SIS} \sim 6.5/\theta$.  
For an isothermal sphere, another image of the background object will
be produced when $\kappa > 0.5$.  This condition is satisfied for the
foreground cluster whether the cluster centroid is estimated using the
luminosity-weighted position or the straight median position.  Using
these centroids leads to $\kappa = 1.1$ or $\kappa = 0.51$,
respectively, at the position of the lens.  We have not seen evidence
for an obvious lensed counterpart in deep 5~GHz or 8.5~GHz radio maps,
although there are other compact sources in the field \citep[e.g.,
object 2;][]{1608delays1}.  Furthermore, such large convergences,
whether from a SIS or some other mass distribution would almost
certainly lead to an image separation in the main lens larger than the
2\farcs1 that is observed.  Once again, these arguments indicate that
Group 4 is not a real cluster.
Because the evidence against the reality of the cluster is strong,
we instead apply the SIS approximation to our arbitrarily selected
subgroups, 4a and 4b.  For these groups, the convergences (Table
\ref{tab_groupconv}) are much more reasonable

The calculation of $\kappa_{ind}$ for Group 4 does not depend on
whether the group is a single cluster or a pair of merging groups or a
filament.  For this group, the galaxy lens strengths are estimated
from the $r$-band magnitudes from the ground-based imaging.  The
fiducial galaxy, which is the brightest confirmed group member, has $r
= 20.0$.  We will discuss different estimates of the lens strength of
the fiducial galaxy below.  We note here, however, that a galaxy at
this redshift will have $b \sim 1$ if its stellar velocity dispersion
is $\sim$260~\kms.

\subsubsection{Convergence Due to Groups 2 and 3}

The convergences calculated for Groups 2 and 3 using the SIS method
are larger than that of Group 1 due to their larger velocity
dispersions.  For Group 3, the two SIS estimates of the convergences
are similar because the group is compact and located approximately
1\arcmin\ from the lensing galaxy.  In contrast, because Group 2 is
roughly centered on the lens system, small displacements in the
centroid can lead to large changes in $\kappa_{SIS}$.  For both Group
2 and Group 3, the fiducial galaxies used in the $\kappa_{ind}$ and
$\kappa_{trunc}$ methods are once again the brightest galaxies in
their respective groups.  In each case the fiducial galaxy has $r =
20.0$.  Galaxies at the redshifts of Groups 2 and 3 have $b \sim 1$
for stellar velocity dispersions of $\sim$220~\kms\ and
$\sim$230~\kms, respectively.

\subsubsection{Total Convergences of Groups 2, 3, and 4}

In order to compute the convergences for Groups 2--4, we need to
assign a value for $b_{fid}$ for each group.  We do this by assuming
that the fiducial galaxy in each group has a velocity dispersion of
200~\kms.  This is slightly smaller than the expected dark matter
velocity dispersion for a $L^\ast$ galaxy of $\sigma^\ast_{DM} \sim
225$~\kms\ \citep[e.g.,][]{csklambda}.  Because the fiducial galaxy in
each case is the brightest one in its respective group, this
assumption should be reasonable.  The resulting lens strengths are
0.85, 0.76, and 0.61
for Groups 2, 3, and 4, respectively.  The resulting convergences are
given as the upper end of the ranges given for $\kappa_{ind}$ and
$\kappa_{trunc}$ in Table~\ref{tab_groupconv}.  These are within a
factor of a few of the convergences estimated from assuming that the
groups were isothermal spheres.

One factor that could lessen the total convergence arises from the
conversion of the group galaxy luminosities to masses.  For example,
in the ACS images, several of the confirmed members of the groups
appear to be late-type galaxies.  In this case the estimated masses
are probably too high since spirals have lower mass-to-light ratios
than ellipticals \citep[e.g.,][]{bld_dm}.  To explore this effect, we
assume that the fiducial galaxy for Groups 2, 3, and 4 has a velocity
dispersion of 140~\kms.  The resulting lens strengths are $\sim$50\%
as large as those obtained above.  The corresponding convergences are
given as the lower ends of the ranges listed in the $\kappa_{ind}$ and
$\kappa_{trunc}$ columns of Table~\ref{tab_groupconv}.

In contrast, it is almost certain that not all of the group members
have been identified.  An incomplete census of the group will lead to
an underestimate of the total convergence.  We note that the deep ACS
imaging of this field reveals at least 10 galaxies within 10\arcsec\
of the lens, corresponding to $\sim 40 - 90~h^{-1}$ comoving kpc at
redshifts between 0.3 and 1.0.  Only one of these galaxies has a
measured redshift ($z = 0.6087$).  Overall, a reasonable estimate is
that true convergence from each group falls between the lower end of
the $\kappa_{trunc}$ range and the largest of the other estimates of
$\kappa_{ext}$.

\subsection{Stellar Velocity Dispersion of the Lensing Galaxy}

Until now we have ignored another datum which can be used to break the
mass-sheet degeneracy, namely the measured stellar velocity dispersion
of the lensing galaxy.  This measurement provides an estimate of the
enclosed mass at a smaller radius than that probed by the lensing
signature.  Thus, the mass measurement from lensing can be combined
with that from stellar dynamics to provide an effective mass density
slope in the lensing galaxy \citep[e.g.,][]{ktlsd,tklsd}.  The result
of adding an external mass sheet that is physically associated with
the lensing galaxy is to flatten the overall mass density profile,
which leads to a lower value of $H_0$ for given time delays.  A
measurement of the stellar velocity dispersion will reflect this
flattening, i.e., the velocity dispersion will not be as high as would
have been predicted from the lensing mass and an assumption of a
steeper mass density profile.  For a density profile
that is close to isothermal (i.e., $\rho \propto r^{-2}$), essentially
the same value of $H_0$ should be derived whether the system is modeled
as a single galaxy with the effective mass density slope measured from
lensing plus dynamics or as a galaxy with a steeper density profile 
but also including an external mass sheet \citep{leonmasssheet}.  For a
high-accuracy measurement of $H_0$, it is thus crucial to obtain high
precision measurements of either the external convergence or the
stellar velocity dispersion of the lensing galaxy.
In the case of B1608+656, 
the measured velocity dispersion is $247 \pm 35$ \kms, indicating that
the mass distribution in the lens (including any contribution from the
group) is close to isothermal \citep{1608H0}.  The uncertainties in the
velocity dispersion are, in fact, the largest source of error in the 
current model.  

\subsection{Effect of Large-scale Structure}

Finally, we consider the effect of the mass in large-scale structures
(LSS) along the line of sight to the background source.  In an ideal
situation, it would be possible to account for all mass along the line
of sight and to trace the rays from the background source through the
distribution.  In practice, even with deep space-based imaging, this
is impractical.  The uncertainties arising from photometric redshifts,
the conversion of light to mass, etc., would far exceed the expected
size of the effect.  Therefore, it is necessary to examine the effect
of LSS in a statistical manner.  Analytic calculations
have indicated that large-scale structure should affect the time
delays for a lens system, and hence $H_0$, by a few percent 
\citep{seljaklss} or perhaps as much as 10\%, depending on the
redshift of the background source \citep{barkanalss}.  
The effect of LSS can be to either increase or decrease the value of
$H_0$ because voids along the line of sight effectively act as areas
of negative density when compared to the mean density of the Universe.
Although this effect is a systematic one for any given lens system, it
should be random for random lines of sight
\citep[e.g.,][]{seljaklss}.  Therefore, it should be possible to
significantly reduce the uncertainty due to large-scale structures on
the global measurement of $H_0$ by averaging the values obtained from
many lens systems.  We note that lenses may lie along lines of sight
that are biased (e.g., have more line-of-sight structure) and
therefore that the effect of large-scale structure can not be
eliminated completely by averaging lens-based measurements of $H_0$.
We will investigate this question in a future paper (Fassnacht et al.,
in prep).

\subsubsection{The Effect on $H_0$}

Finally, we arrive at the total effect on the value of $H_0$ derived
from the B1608+656 system.  The group physically associated with the
lens system, Group 1, provides an external convergence of $\sim 0.01 -
0.03$ (Table~\ref{tab_groupconv}).  As discussed above, however, the
degeneracy associated with Group 1 is broken by the measurement of the
stellar velocity dispersion of the lensing galaxy.  Therefore, the
effect of Group 1 on the value of $H_0$ derived from this system gets
folded into the determination of the radial mass slope in the lensing
galaxy, and is thus already included in the published uncertainties on
$H_0$ from B1608+656.

The other groups along the line of sight also provide convergences of
a few percent.  These groups are almost certainly typical of the
clumpy structure that can be found along any line of sight.  None of
them provide an extraordinarily large convergence, unless Group 4 is a
real cluster.  Therefore, one would be inclined to incorporate the
effects of these groups into the effects of LSS and conclude that
these groups will contribute to the random shift of a few percent in
the value of $H_0$ determined from this system.  However, as a result
of our investigation of lens fields (Fassnacht et al., in prep) we
have determined that the B1608+656 system lies along a line of sight
that is overdense compared to typical lines of sight.  Therefore, the
expected effect of properly incorporating the effects of LSS on this
system is to {\em reduce} the value of $H_0$ from the previously
measured value.  If the majority of the extra mass along this line of
sight is being contributed by Groups 1-3, the size of the effect could
be 5\% or more, given the convergences in Table~\ref{tab_groupconv}.
This effect, thus, could be comparable in size to the statistical
uncertainties quoted for the $H_0$ measurement from this system, and
would reduce the central value to $\sim$70~\kmsmpc\ or lower.

\section{Summary and Future Work
 \label{summary}
}

Our spectroscopic observations of the field containing the lens system
B1608+656 have provided evidence for four groups of galaxies along the
line of sight to the lens system.  These groups should contribute
external convergence at the location of the lens and, therefore, will
affect the determination of the Hubble Constant obtained with this
system.  However, quantifying the amount of external convergence is
difficult.  If each group contains all of its mass in a overall
smooth halo, it becomes imperative to measure accurately the halo mass
and centroid.  These measurements can be highly uncertain when made
based on $\sim 10$ redshifts.  We therefore follow a second approach
and treat each group as a collection of individual galaxy halos.  To
establish a firm lower limit to the convergence due to each group, we
examine the convergence contributed only by galaxies within a
projected distance of 200~$h^{-1}$ comoving kpc from the lens.
However, our calculations do not include contributions from other
galaxies within the truncation radius of the lens because their
redshifts are unknown.  Each of the groups contributes a convergence
of a few percent.  The overall effect on $H_0$ is less than suggested
by the sum of the convergences because the stellar velocity dispersion of the
main lensing galaxy has been measured.  This measurement breaks the
mass-sheet degeneracy, at least due to Group 1, by determining the
effective radial mass density slope in the lensing galaxy.  The effects of
the other groups can be folded into the overall uncertainties due to
large-scale structure along the line of sight, since none of the
groups appears to be extremely massive.  However, because the line
of sight to B1608+656 appears to be significantly overdense compared
to typical lines of sight, the effect of LSS will be to bias the
current measurement of $H_0$ high, i.e., the true value of $H_0$ from
this system should be lower than the published value.  The size of the
effect of the LSS could be 5\% or larger.
We note that none of the newly discovered groups was
obvious from optical images of this system.  Therefore, it is
important to closely investigate the fields of lens systems in order
to examine possible sources of bias in lens-based measurements of $H_0$,
especially if the stellar velocity dispersion of the lensing galaxy
is unknown.

We are actively working to reduce the uncertainties in the
determination of $H_0$ from this lens system.  The deep ACS imaging is
being used as an input to new lensing code that can properly
incorporate Einstein ring structure.  This modeling should reduce the
mass-slope uncertainties substantially in the region of the lensed
images.  We will also use the ACS imaging to search for weak lensing
signatures of mass concentrations along the line of sight to the lens
system.  The weak lensing will provide further constraints on the
amount of external convergence for B1608+656.  Another approach to
reducing the uncertainties in the mass slope would be to obtain higher
sensitivity spectroscopy of the lensing galaxy.  The spectroscopy
would provide a more accurate determination of the stellar velocity
dispersion.  Further optical spectroscopy of the galaxies in the field
and deeper X-ray imaging would also add information on the external
mass distribution.  Finally, by averaging measurements of $H_0$
obtained from many different lens systems, we should be able to reduce
the uncertainties due to large-scale structure and thus obtain a
precise global measurement of $H_0$ from lensing.


\acknowledgments 

We thank Leon Koopmans, Tommaso Treu, and Tony Tyson for useful
discussions.  We thank the anonymous referee for his or her comments.
CDF and JPM acknowledge support under HST program \#GO-10158.
Support for program \#GO-10158 was provided by NASA through a grant from
the Space Telescope Science Institute, which is operated by the
Association of Universities for Research in Astronomy, Inc., under
NASA contract NAS 5-26555.  
These observations would not have been possible without the expertise
and dedication of the staffs of the Palomar and Keck observatories.
We especially thank Paola Amico, Karl Dunscombe, Grant Hill, Jean
Mueller, Ron Quick, Kevin Rykoski, Gabrelle Saurage, Chuck Sorenson,
Skip Staples, Wayne Wack, Cindy Wilburn, and Greg Wirth.
Some of the data presented herein were obtained at the W. M. Keck
Observatory, which is operated as a scientific partnership among the
California Institute of Technology, the University of California and
the National Aeronautics and Space Administration. The Observatory was
made possible by the generous financial support of the W.M. Keck
Foundation.  The authors wish to recognize and acknowledge the very
significant cultural role and reverence that the summit of Mauna Kea
has always had within the indigenous Hawaiian community.  We are most
fortunate to have the opportunity to conduct observations from this
mountain.
This work is supported in part by the European Community's Sixth
Framework Marie Curie Research Training Network Programme, Contract
No.\ MRTN-CT-2004-505183 `ANGLES'.


\ifsubmode
   {\it Facility:} \facility{HST (ACS, WFPC2)}, \facility{Keck:I (LRIS)}, 
   \facility{Keck:II (ESI)}, \facility{PO:1.5m (CCD13)}
\fi




\end{document}